\documentclass{aastex}
\usepackage{emulateapj5}
\begin{document}

\title{A Period and a Prediction for the Of?p Spectrum Alternator HD~191612}

\author{Nolan~R.~Walborn\altaffilmark{1},
        Ian~D.~Howarth\altaffilmark{2,3},
        Gregor~Rauw\altaffilmark{4,5},
        Daniel~J.~Lennon\altaffilmark{6},
        Howard~E.~Bond\altaffilmark{1,7},
        Ignacio~Negueruela\altaffilmark{8,9},
        Ya\"el~Naz\'e\altaffilmark{4},
        Michael~F.~Corcoran\altaffilmark{10},
        Artemio~Herrero\altaffilmark{11},
    and Anne~Pellerin\altaffilmark{12}}

\altaffiltext{1}{Space Telescope Science Institute, 3700 San Martin Drive, 
Baltimore, MD 21218.  STScI is operated by the Association of Universities 
for Research in Astronomy, Inc., under NASA contract NAS5-26555;
walborn@stsci.edu, bond@stsci.edu.}

\altaffiltext{2}{Department of Physics and Astronomy, University College 
London, Gower Street, London WC1E 6BT, UK;
idh@star.ucl.ac.uk.}

\altaffiltext{3}{Visiting Observer, Isaac Newton Group, La Palma, Spain.}

\altaffiltext{4}{Institut d'Astrophysique et de G\'eophysique, Universit\'e 
de Li\`ege, All\'ee du 6 Ao\^ut 17, B\^at B5c, B4000 Li\`ege, Belgium.
Also at the Fonds National de Recherche Scientifique, Belgium;
rauw@astro.ulg.ac.be, naze@astro.ulg.ac.be.}

\altaffiltext{5}{Visiting Observer, Observatoire de Haute-Provence, France.
The 2004 OHP run was funded by the OPTICON network supported under the FP6 
program of the European Commission.}

\altaffiltext{6}{Isaac Newton Group, Apartado 321, 38700 Santa Cruz de 
La Palma, Canary Islands, Spain;
djl@ing.iac.es.}

\altaffiltext{7}{Visiting Astronomer, Kitt Peak National Observatory, 
National Optical Astronomy Observatory, operated by AURA, Inc. under 
cooperative agreement with the NSF; and Multiple-Mirror Observatory, 
University of Arizona.}

\altaffiltext{8}{Departamento de F\'{\i}sica, Ingenier\'{\i}a de Sistemas
y Teor\'{\i}a de la Se\~nal, Universidad de Alicante, Apartado 99, E03080
Alicante, Spain.  Researcher of the {\em Ram\'on y Cajal} program, funded
by the Spanish Ministerio de Ciencia y Tecnolog\'{\i}a and the University 
of Alicante.  This research is partially supported by the Spanish MCyT under 
grant AYA2002-00814;
ignacio@dfists.ua.es.}

\altaffiltext{9}{Visiting Observer, Loiano Observatory, Italy.}

\altaffiltext{10}{Universities Space Research Association, 7501 Forbes 
Boulevard, Suite 206, Seabrook, MD 20706 and Laboratory for High Energy 
Astrophysics, Goddard Space Flight Center, Greenbelt, MD 20771;
corcoran@barnegat.gsfc.nasa.gov.}

\altaffiltext{11}{Instituto de Astrof\'{\i}sica de Canarias, 38200 La Laguna,
Tenerife, Spain;
ahd@ll.iac.es.}

\altaffiltext{12}{D\'epartement de physique, de g\'enie physique et d'optique 
and Observatoire du mont M\'egantic, Universit\'e Laval, Qu\'ebec, QC, G1K 7P4, 
Canada;
anpel@phy.ulaval.ca.}

\begin{abstract}
The observational picture of the enigmatic O-type spectrum variable HD~191612 
has been sharpened substantially.  A symmetrical, low-amplitude light curve 
with a period near 540~d has recently been reported from {\it Hipparcos} 
photometry.  This period satisfies all of the spectroscopy since at least 
1982, including extensive new observations during 2003 and 2004, and it has
predicted the next transition during September--October 2004.  Measurements 
of the H$\alpha$ equivalent width reveal a sharp emission peak in the phase 
diagram, in contrast to the apparently sinusoidal light curve.  The He~II
absorption-line strength is essentially constant, while He~I varies strongly, 
possibly filled in by emission in the O6 state, thus producing the apparent
spectral-type variations.  The O8 state appears to be the ``normal'' one.  
Two intermediate O7 observations have been obtained, which fall at the 
expected phases, but these are the only modern observations of the transitions 
so far.  The period is too long for rotation or pulsation; although there is 
no direct evidence as yet for a companion, a model in which tidally induced 
oscillations drive an enhanced wind near periastron of an eccentric orbit 
appears promising.  Further observations during the now predictable 
transitions may provide a critical test.  Ultraviolet and X-ray observations 
during both states will likely also prove illuminating.
\end{abstract}

\keywords{stars: early-type---stars: emission-line, Be---stars: individual 
(HD~191612)---stars: mass loss---stars: variables: other}

\section{Introduction}

The discovery of two recurrent spectral states in the Of?p star HD~191612 
was described by Walborn et al. (2003).  The spectral type changes between 
O6 and O8, with correlated variations in several peculiar line profiles, 
including the transformation of H$\alpha$ from a strong P~Cygni profile to 
predominantly absorption, respectively.  That paper left the star in the O8 
state at the end of 2002, with considerable uncertainty about the (long) 
timescale and regularity of the spectral transitions, due to the 
fragmentary observational record.  This Letter updates the observational 
situation since then, including the definition of the timescale 
and period; the extensive discussions of the history and possible 
interpretations of the phenomenon in the earlier paper will not be repeated 
here.  The most significant new development is that the transition epochs
can now be predicted, so that intensive observations during them can be
planned.  This development may well enable progress toward an
interpretation of the bizarre phenomenon, the characteristics of which
are unprecedented in any O-type star.

As will be shown below, in May 2003 HD~191612 was found to have returned
to the O6 state, where it remained through October of that year, but a
December observation showed signs of a transition, and it was recovered in 
the O8 state in May 2004.  Thus, spectral states last less than a year, most 
likely 7-9 months based on the spectroscopy alone.  Concurrently, Naz\'e 
(2004; see also Koen \& Eyer 2002) reported a symmetrical, very low-amplitude 
light curve from {\it Hipparcos} photometry of this star, with a period of 
about 536 days or 18 months.  This period is consistent with the timescale 
indicated independently by the spectroscopy.  We rederived the period
from a sine-wave fit to the {\it Hipparcos} data weighted by their errors, 
which yielded a value slightly larger than 540~d with an uncertainty of 
$\pm 13$~d.  On the assumption that the period is stable over the longer 
baseline provided by the spectroscopic observation of Peppel (1984), we
then refined its value to $538 \pm 3$~d, where the stated uncertainty moves 
that point by 0.08 in phase, displacing it significantly from the trend.  
With this period we are thus able to phase all spectroscopic observations 
since 1982, deriving tight correlations among the spectral variations.  It 
even accommodates the spectral type of Walborn (1973), indicating that the 
period may be accurate and stable to at least a decade earlier.  These results 
provide a strong prediction of the next transition (from the current O8 to 
the O6 state) during October 2004, with the H$\alpha$ emission already rising 
significantly during September.

\section{New Observations}

The 2003 and 2004 observations to date are listed in Table~1 and most of 
them are displayed in Figures~1 (blue-green wavelengths) and 2 (yellow-red).  
Fifteen epochs (some containing multiple observations) were obtained between 
May and December 2003.  The first of these showed that the star had (nearly) 
returned to the O6 state since the last 2002 observation (also shown for 
reference in Fig.~1; see Walborn et al. 2003).  There is evidence of the 
transition in the May~15 spectrum, in the almost equal He~I $\lambda4471$ 
and He~II $\lambda4541$ absorption-line strengths (corresponding to spectral 
type O7), and in the smaller ratio of the C~III $\lambda4650$ to N~III 
$\lambda4640$ emission-line strengths than subsequently.  The December~6 
observation is remarkably similar, with the addition of an intermediate 
H$\alpha$ emission strength, indicating the onset of another transition, and 
that the complete duration of the O6 state was covered during 2003.  The 2003 
data provide several additional new results with respect to those available to 
Walborn et al. (2003).  They are the first observations of the O6 state since 
the variability was discovered and include the first simultaneous blue and 
red digital observations in the O6 state, confirming inferences about their
relationship in the earlier paper.  A P~Cygni profile is observed for the
first time in He~I $\lambda5876$, and the composite H$\beta$ profile aids 
the interpretation of the H$\alpha$ and H$\gamma$ profiles.  Indeed, it
is now clear from a comparison of the behavior of the three hydrogen lines 
that the peculiar profile of the H$\alpha$ absorption in the O8 state is
caused by the persistence of a weak P~Cygni profile in its core.  In the
O6 state, emission is visible in the Balmer lines to H$\zeta$.

We have observed at ten epochs in May-July 2004, which document a new
O8 state and definitely establish that the timescale of the spectral
variations is less than a year.  Thus, the possibility of a decade
timescale allowed by the data available to Walborn et al. (2003) is ruled
out, and evidence for the shorter timescale suggested by trends within
the 2002 data is confirmed.  As noted above, this shorter timescale is in
excellent agreement with the independent photometric period found in the
{\it Hipparcos} data.

\section{Discussion}

\subsection{Synthesis and Prediction}

Figure~3 displays the {\it Hipparcos} phase diagram together with 
measurements of key hydrogen and helium spectral features.  The 
{\it Hipparcos} observations were done between November 1989 and February 
1993, while the digital spectroscopic observations extend from June 1989 
through July 2004.  Blind remeasurements show that the equivalent widths 
typically repeat to better than 0.1~\AA, which is consistent with the 
observed scatter about the trends; the H$\alpha$ point with large error bar 
is a visual estimate from the plot published by Peppel (1984), corresponding 
to an observation on August 29, 1982, which has been used to refine the value 
of the period as discussed above.  The individual phases and measurements
of these and other features will be listed in a subsequent paper.  The tight 
correlations among the spectral features across 15 cycles demonstrate that 
the derived period is sufficiently accurate to phase them properly.  
Surprisingly, the H$\alpha$ P~Cygni emission has a sharp peak, in contrast 
to the apparently smooth light curve.  As shown, the transition from O8 to
O6, which has been practically unobserved by the prior spectroscopy, will
next occur centered in October 2004, while a significant increase in the
H$\alpha$ emission will begin during September.

Another significant conclusion from Fig.~3 is that the O6 state (He~II
stronger than He~I) occurs at maximum light, and conversely the O8 state
at minimum.  This is the opposite relationship to that predicted in
Walborn et al. (2003), on the assumption of constant bolometric
luminosity.  The He~II line displays little if any systematic variation 
in Fig.~3; in fact, little variation is predicted between 35,000~K and
40,000~K at the relevant gravity and mass-loss rates.  However, it is
also possible that the stellar effective temperature is not changing, and 
that the weaker He~I in the O6 state is due to emission filling of the
absorption line.  Thus, the spectral-type variation may be only apparent.
Clearly, better spectroscopic coverage of the transition phases will be
essential to delineate the full behavior of the diagnostic features.  

\subsection{Physical Constraints}

The characteristics of the HD~191612 variability present serious
difficulties for all of the usual mechanisms in early-type stars.
The period of $\sim540$~d is far too long to be rotational.  A $v$sin$i$ of
77~km~s$^{-1}$ was derived in Walborn et al. (2003), which is an
underestimate of the equatorial rotational velocity because of the
unknown inclination, but may be an overestimate if turbulence or a 
velocity gradient contributes to the line broadening.  An (average)
radius of 17.7~$R_{\sun}$ was also derived in that paper, which for a
fiducial rotational velocity of 100~km~s$^{-1}$ corresponds to a period
of 9~d.  This issue also rules out any straightforward magnetic
hypothesis for the variations as in an oblique rotator (although it may
be well to recall that the poorly understood solar cycle is far longer 
than the rotational period of the Sun), and it is similarly adverse to
any binary model involving synchronous rotation.  A precession of the
rotational axis of a distorted star with different polar and equatorial
spectra and winds might be considered, but then the apparent rotational 
velocity should also undergo extreme variations, and the physical basis 
of such a mechanism is unclear at best.

The period is also too long for a pulsational timescale in an O star: the 
sound-crossing time for a star of 40~$M_{\sun}$ and 20~$R_{\sun}$ yields a 
radial-mode period of $\sim$10~h, and the nonradial pulsations observed
in a few O stars are also of the order of hours.  A beat period of two 
pulsational modes might be considered, however, as has been applied to 
explain variations in the mass-loss rate and spectrum of the Be star 
$\mu$~Centauri (Rivinius et al. 2001).  There is no direct observational 
evidence for pulsations in HD 191612 at present. 

An upper limit of a few km~s$^{-1}$ to any radial-velocity variations in
HD~191612 was derived in Walborn et al. (2003), which is not inconsistent 
with a low-mass companion in a 540~d orbit.  For instance, a 2~$M_{\sun}$ 
companion in a circular orbit would induce a semi-amplitude of only 
4.3~km~s$^{-1}$ at $90^{\circ}$ inclination.  However, such a system would 
not provide sufficient tidal distortion to induce large spectral 
variations in the primary, and the problem of the synchronous rotation 
required for phase-locked variations has already been noted.  An eccentric 
orbit with a small periastron separation is not ruled out by the 
radial-velocity data and might be consistent with the H$\alpha$ behavior,
although the symmetrical light curve would then be a puzzle.  An
interesting prospect in this regard is the model of tidally induced
nonradial pulsations at periastron, in a system with nonsynchronized
rotational and orbital angular velocities (Koenigsberger, Moreno, \&
Cervantes 2002, 2003; Willems \& Aerts 2002).  Such pulsations could in turn
drive enhanced mass loss (Owocki \& Cranmer 2002), consistent with the
spectral variations in HD~191612.  The Herbig Be system HD~200775 may be
an example (Pogodin et al. 2004).  Another alternative in principle 
might be X-ray irradiation from a collapsed companion as a 
spectral-variability mechanism, but HD~191612 is a relatively weak source, 
and the much stronger source LMC X-4 with a 1.4~d period produces 
only small variations in the spectrum of the O-type primary (Negueruela \& 
Coe 2002 and references therein).  We note that a short rotational period 
would not be a problem in this case, however, if the response time of the 
stellar atmosphere were much less.  Chlebowski (1989) suggested that the 
Of?p stars might be binaries with collapsed companions, but Naz\'e et al. 
(2004) concluded that HD~191612's Of?p spectral classmate HD~108 is unlikely 
to be an X-ray binary from {\it XMM-Newton} observations; HD~108 displays the
same kinds of optical spectral variations as HD~191612, but on a timescale of
decades (Naz\'e, Vreux, \& Rauw 2001).  Although we shall perform 
radial-velocity measurements in our new data, we can rule out any variations 
larger than 10-20~km~s$^{-1}$, and we are not optimistic about the precision 
with which smaller values can be measured, because many of the data are not 
optimized for that purpose, and the systematic line-profile differences 
between the two spectral states will obscure any small radial-velocity 
variations.  However, if the enhanced wind is a periastron event, the most 
favorable phases for the detection of radial-velocity variations have not yet 
been well observed (Figure~3).  The observed period is too short for an 
orbital precession, as found in the hierarchical triple system IU Aurigae by 
Drechsel et al. (1994).

Perhaps the most intriguing hypothesis is that the strange behavior of
HD~191612 might be related to stellar evolution, via the Eddington Limit
and/or the Luminous Blue Variable phenomenon; a binary connection 
to the latter is by no means ruled out.  The other Galactic Of?p spectral
classmate HD~148937 has ejected a spectacular, axisymmetric, nitrogen-rich
nebula (Walborn et al. 2003 and references therein).  Could HD~191612 be 
approaching a similar event?  The short timescale and periodicity of the 
observed phenomena would be surprising unless a binary connection is indeed 
involved.  Current understanding of such events is extremely limited, and this 
star could be providing critical new information about the outburst mechanism.

\subsection{Outlook}

The new results presented in this Letter have transformed the variations
of HD~191612 from a temporally undefined, apparently random phenomenon
into one that can be predicted.  The observational implications are 
substantial: instead of undirected monitoring, intensive campaigns at the 
key transitional and extreme epochs can be planned; we plan such for Fall 
2004.  Clearly it will be of great interest to obtain higher quality X-ray
observations, as well as observations of the ultraviolet wind profiles,
during both spectral states of HD~191612.  The $IUE$ observation shown by
Walborn et al. (2003) was obtained at phase 0.23, i.e. during a transition.  
If the entire wind is changing, as suggested by the H$\alpha$ variations, 
very large effects in the UV wind profiles would be expected, so observations 
at the peaks of the two spectral states are of high interest. 
It may be hoped that this new information will provide critical guidance 
toward a viable physical model of the phenomenon; despite all indications to 
the contrary, we remain confident that there must be one!  Perhaps the Of?p
class is a mere curiosity among the OB Zoo (Walborn \& Fitzpatrick 2000), 
but it is also possible that it may be offering a missing link in massive 
(binary?) stellar evolution.

\acknowledgments
We are grateful to the additional observers listed in Table~1 who
contributed to our 2003 and 2004 datasets.  Publication support was
provided by the STScI Director's Discretionary Research Fund.

\noindent {\it Note in Press}: Several spectroscopic observations during 
September and October 2004, obtained at WHT, TNG, NOT (La Palma); OHP; and 
Skinakas (Crete), show that the transition is in progress exactly as predicted. 
Measurements of H$\alpha$ and the He lines fall precisely on the trends 
in Figure~3.  These results and measurements from all available data will 
be discussed in a forthcoming paper.

\begin{deluxetable}{lllcc}
\tablewidth{0pt}
\tablecolumns{5}
\tablecaption{New Spectroscopic Observations of HD 191612}
\tablehead{
\colhead{UT Date} &
\colhead{Telescope} &
\colhead{Observer}  &
\colhead{Band} &
\colhead{Res (\AA)}}
\startdata
2003\\
May 15       &WIYN        &H. Bond         &B     &0.6\\
June 11      &INT         &D. Lennon       &BYR   &1.2, 0.6\\
June 13      &INT         &D. Christian    &B     &0.3\\
June 18      &WHT         &A. Herrero      &BYR   &0.7, 0.8\\
June 19      &WIYN        &H. Bond         &B     &0.6\\
June 26      &INT         &D. Lennon       &BYR   &0.5\\
July 11      &INT         &F. Prada        &BR    &0.8, 0.7\\
July 11--18  &INT         &I. Howarth      &BYR   &1.4, 1.1\\
July 24      &WIYN        &D. Harmer       &B     &0.6\\
Aug 2        &WIYN        &D. Harmer       &B     &0.6\\
Aug 19--21   &OMM         &A. Pellerin     &BYR   &3.9\\
Sept 19      &WIYN        &H. Bond         &B     &0.6\\
Oct 1        &MMT         &H. Bond         &BY    &3.6\\
Oct 4--9     &OHP         &G. Rauw         &BR    &0.6\\
Dec 6        &WHT         &D. Lennon       &BYR   &1.8, 1.9\\
\\  
2004\\
May 2        &OHP         &G. Rauw         &BYR   &0.1\rlap{2}\\
May 6        &WIYN        &D. Harmer       &B     &0.6\\
May 21       &WHT         &J. Licandro     &BYR   &1.6, 1.7\\
May 21       &Skinakas    &P. Reig         &BYR   &2.0\\
June 1       &WHT         &I. Howarth      &BR    &0.8\\
June 23--25  &Skinakas    &P. Reig         &BYR   &2.0\\
July 7--8    &Skinakas    &P. Reig         &YR    &2.0\\
July 15      &Loiano      &I. Negueruela   &BR    &3.9, 3.0\\
July 16      &Loiano      &I. Negueruela   &BYR   &1.2\\
July 23      &WHT         &R.C. Smith      &YR    &0.7\\
\enddata
\end{deluxetable}
\clearpage

\begin{figure}
\centerline{\epsfysize=8in\epsfbox{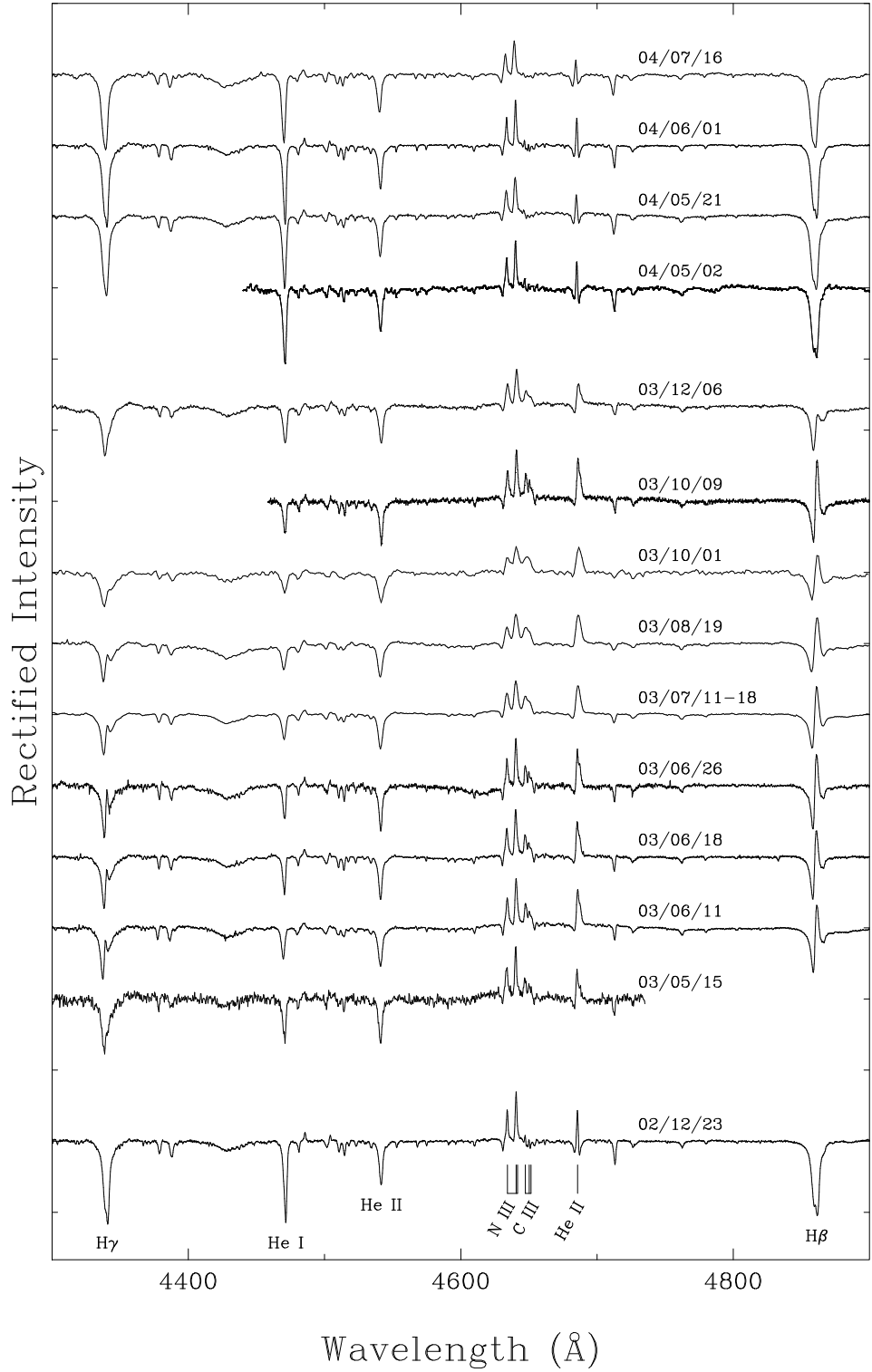}}
\caption{Rectified blue-green spectrograms of HD~191612 during 2003 and
2004; the last 2002 observation from Walborn et al. (2003) is also
plotted.  The ordinate ticks represent 0.3 continuum units.  The spectral
features identified below are H$\gamma$ $\lambda$4340 and H$\beta$ 
$\lambda$4861; He~I $\lambda$4471; He~II $\lambda\lambda$4541, 4686;
N~III $\lambda\lambda$4634-4640-4642; and C~III $\lambda\lambda$4647-4650-4651.
The 2002 and 2004 spectra correspond to the O8 state, while the 2003 show
the O6 state except that the May 15 and Dec. 6 are transitional.
\label{fig1}}
\end{figure}
\clearpage
\begin{figure}
\centerline{\epsfysize=8in\epsfbox{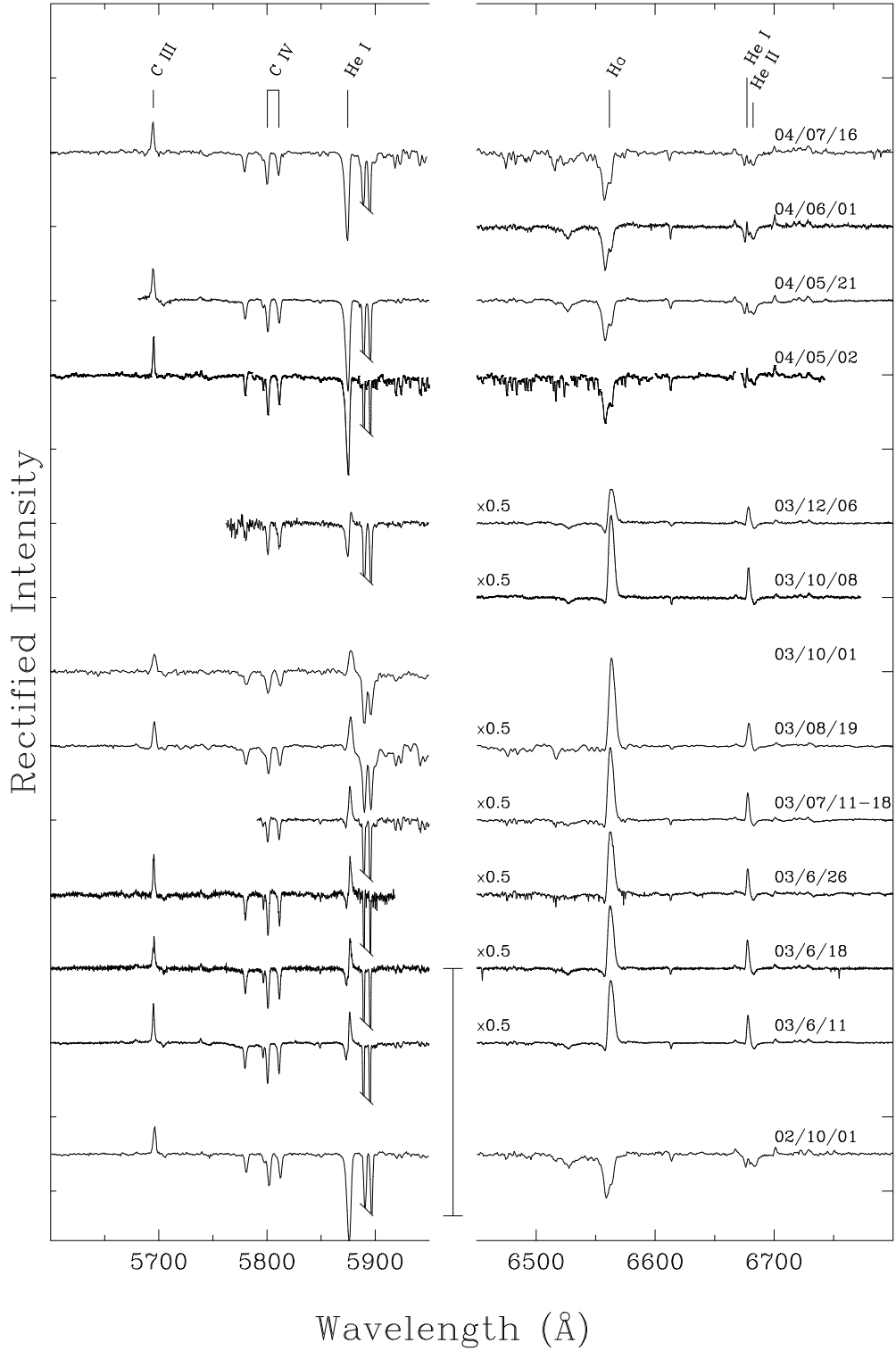}}
\caption{Rectified yellow and red segments of the spectrum of HD~191612
during 2003 and 2004; a late 2002 observation from Walborn et al. (2003) is 
also plotted. The vertical line segment at lower center denotes the
continuum scale, except that the 2003 red spectra with H$\alpha$ emission
have been reduced by a factor of 0.5.  The interstellar Na~I D lines have
been truncated in the higher resolution data.  The spectral features 
identified above are C~III $\lambda$5696; C~IV $\lambda\lambda$5801-5812; 
He~I $\lambda\lambda$5876, 6678; H$\alpha$ $\lambda$6563; and He~II 
$\lambda$6683.  The 2002 and 2004 spectra correspond to the O8 state, while 
the 2003 show the O6 state except that the Dec. 6 is transitional.
\label{fig2}}
\end{figure}
\clearpage
\begin{figure}
\centerline{\epsfysize=8in\epsfbox{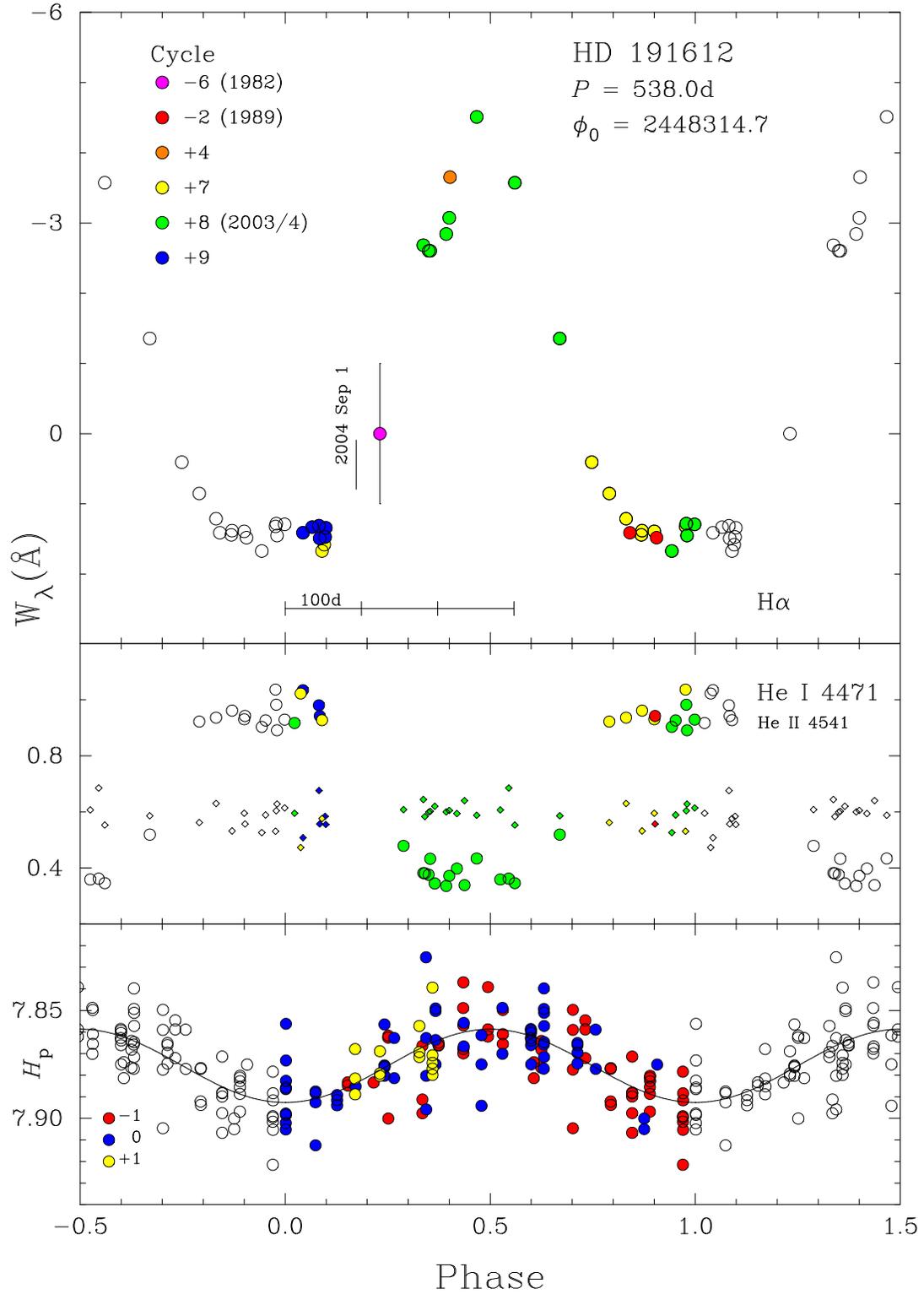}}
\caption{The phased {\it Hipparcos} light curve of HD~191612 ({\it bottom});
phase zero corresponds to the light-curve minimum.  The sine-wave fit yields 
a mean magnitude of $7.876 \pm 0.001$ and an amplitude of $0.034 \pm 0.003$.  
Equivalent-width measurements of diagnostic He~I ({\it large circles}) and 
He~II ({\it small diamonds}) absorption lines ({\it center}) show that the 
apparent spectral type is earlier when the star is brighter, but the variation 
is entirely due to the He~I line.  The H$\alpha$ P~Cygni emission ({\it top}) 
is sharply peaked near phase 0.5, in contrast to the smooth light curve.  As 
shown, the O8 to O6 transition, which has not been covered by the prior 
digital spectroscopy, will next occur during September--October 2004.  
\label{fig3}}
\end{figure}

\end{document}